\documentclass[11pt,english]{article}
\usepackage{rotating}
\usepackage{latexsym}
\usepackage{graphicx}
\usepackage{psfrag}
\usepackage{amsmath,amssymb}
\makeatletter
\def\section{\@startsection {section}{1}{\z@}{-3.5ex plus -1ex minus
 -.2ex}{2.3ex plus .2ex}{\large\bf}}
\def\subsection{\@startsection{subsection}{2}{\z@}{-3.25ex plus -1ex
minus -.2ex}{1.5ex plus .2ex}{\normalsize\bf}}
\makeatother
\makeatletter

\@addtoreset{equation}{section}

\makeatother

\newcommand{\captionfonts}{\small}
\makeatletter  
\long\def\@makecaption#1#2{%
  \vskip\abovecaptionskip
  \sbox\@tempboxa{{\captionfonts #1: #2}}%
  \ifdim \wd\@tempboxa >\hsize
    {\captionfonts #1: #2\par}
  \else
    \hbox to\hsize{\hfil\box\@tempboxa\hfil}%
  \fi
  \vskip\belowcaptionskip}
\makeatother   
\def\dslash{\raisebox{1pt}{$\slash$} \hspace{-6pt} \partial}
\def\pslash{\raisebox{1pt}{$\slash$} \hspace{-8pt} p}

\def\Aslash{\hspace{3pt}\raisebox{1pt}{$\slash$} \hspace{-9pt} A}
\def\Dslash{\hspace{3pt}\raisebox{1pt}{$\slash$} \hspace{-8pt} D}

\def\bea{\begin{eqnarray}} \def\eea{\end{eqnarray}}
\def\be{\begin{equation}} \def\ee{\end{equation}} \def\nn{\nonumber}
\def\a{& \hspace{-7pt}}  \def\Z{{\bf Z}}
 \def\ov{\overline}

\def\mc{\mathcal}

\newcommand{\promille}{%
  \relax\ifmmode\promillezeichen
        \else\leavevmode\(\mathsurround=0pt\promillezeichen\)\fi}
\newcommand{\promillezeichen}{%
  \kern-.05em%
  \raise.5ex\hbox{\the\scriptfont0 0}%
  \kern-.15em/\kern-.15em%
  \lower.25ex\hbox{\the\scriptfont0 00}}

\begin{document}

\thispagestyle{empty}

\begin{center}
\hfill UAB-FT-602 \\
\hfill SISSA-32/2006/EP \\

\begin{center}

\vspace{1.7cm}

{\LARGE\bf Electroweak Symmetry Breaking and\\[3mm]
Precision Tests with a Fifth Dimension}

\end{center}

\vspace{1.4cm}

{\bf Giuliano Panico$^{a}$, Marco Serone$^{a}$ and Andrea Wulzer$^{b}$}\\

\vspace{1.2cm}

${}^a\!\!$
{\em ISAS-SISSA and INFN, Via Beirut 2-4, I-34013 Trieste, Italy}

\vspace{.3cm}

${}^b\!\!$
{\em { IFAE, Universitat Aut{\`o}noma de Barcelona,
08193 Bellaterra, Barcelona}}

\end{center}

\vspace{0.8cm}

\centerline{\bf Abstract}
\vspace{2 mm}
\begin{quote}\small

We perform a complete study of flavour and CP conserving electroweak observables in
a slight refinement of a recently proposed five--dimensional model
on $R^4\times S^1/\Z_2$, where the Higgs is the internal component
of a gauge field and the Lorentz symmetry is broken in the fifth dimension.

Interestingly enough, the relevant corrections to the electroweak
observables turn out to be of universal type and 
essentially depend only on the value of the Higgs mass
and on the scale of new physics, in our case the compactification scale $1/R$.
The model passes all constraints for $1/R\geq 4.7$ TeV at 90$\%$ C.L., with a moderate
fine--tuning in the parameters.
The Higgs mass turns out to be always smaller than 200 GeV although higher values
would be allowed, due to a large correction to the $T$ parameter.
The lightest non-SM states in the model are typically colored fermions with a mass
of order $1-2$ TeV.

\end{quote}

\vfill

\newpage

\section{Introduction}

The idea of identifying the Higgs field with the internal component of
a gauge field in TeV--sized extra dimensions 
\cite{Antoniadis:1990ew}, recently named Gauge-Higgs Unification (GHU), has received a revival of interest
in the last years. Although not new (see {\it{e.g.}} \cite{early}), only recently it has been realized that
this idea could lead to a solution of the Standard Model (SM) instability of the electroweak scale \cite{GHU}.

Five-dimensional (5D) models, with one extra dimension only, are the
simplest and phenomenologically more appealing examples.
When the extra dimension is a flat $S^1/\Z_2$ orbifold, however, the Higgs and
top masses and the scale of new 
physics, typically given by the inverse of the compactification radius
$1/R$, are too low in the simplest implementations of the
GHU idea \cite{Scrucca:2003ra}. Roughly speaking, 
the Higgs mass is generally too light because the Higgs effective potential
is radiatively induced, giving rise to a too small effective SM-like quartic coupling.  
The top is too light because the Yukawa couplings, which are effective couplings in these theories, are
engineered in such a way that they are always smaller than the electroweak gauge couplings, essentially
due to the 5D Lorentz symmetry.
Finally, the scale of new physics is too low, because in minimal models there is no way to generate
a sizable gap between the electroweak (EW) scale and $1/R$.

It has recently been pointed out that it is possible to get rid of all the above problems by
increasing the Yukawa (gauge) couplings of the Higgs field with the fermions, by
assuming a Lorentz symmetry breaking in the fifth direction \cite{Panico:2005dh} (see also \cite{Agashe:2004rs}
for other interesting possibilities in a warped scenario).
In this way, the Yukawa couplings are not constrained anymore to be
smaller than the electroweak gauge couplings and we have the
possibility to get Yukawa's of order one, as needed for the top quark.
Stronger Yukawa couplings lead also to a larger effective Higgs quartic 
coupling, resulting in Higgs masses above the current
experimental bound of $m_H> 115$ GeV. Furthermore, again due to the larger Yukawa couplings, the
Higgs effective potential is totally dominated by the fermion
contribution. The latter effect, together with a proper choice 
of the spectrum of (periodic and antiperiodic) 5D fermion fields, 
can lead to a substantial gap between the EW and compactification scales.
In non--universal models of extra dimensions of this sort, where the
SM fermions have sizable tree--level couplings with Kaluza--Klein (KK) gauge fields,
the electroweak precision tests (EWPT) generically imply quite
severe bounds on the compactification scale. At the price of some
fine-tuning  on the microscopic parameters of the theory, however, one can push $1/R$ in the multi-TeV
regime, hoping to get in this way a potentially realistic model.

The SM fermions turn out to be almost localized fields, with the exception of the bottom quark, which shows
a partial delocalization, and of the top quark, which is essentially delocalized. This effect
generally leads to sizable deviations of the $Z \overline b_L b_L$
coupling with respect to the SM one. Another distortion is introduced in the gauge sector,
in order to get the correct weak mixing angle (see \cite{Grzadkowski:2006tp} for a recent discussion on the difficulty
of automatically getting the correct weak-mixing angle in 5D GHU models). The main worrisome effect
of the distortion is the departure \cite{Panico:2005dh} of the
SM $\rho$ parameter from one at tree-level.

Aim of this paper is to consider a variant of the model constructed in
\cite{Panico:2005dh} and to compute the flavour conserving
phenomenological bounds. The main new ingredient is an
exact discrete $\Z_2$ symmetry, that we call ``mirror symmetry''. It essentially consists in doubling
a subset of bulk fields $\phi$, namely all the fermions and some of the gauge fields, 
in pairs $\phi_1$ and $\phi_2$ and requiring a symmetry under the
interchange $\phi_1\leftrightarrow \phi_2$. 
Periodic and antiperiodic
fields arise as suitable linear combinations of $\phi_1$ and $\phi_2$
and the symmetry constrains the couplings of the Higgs
with periodic and antiperiodic fermions with the same quantum
numbers to be equal. When this is not the case, as in \cite{Panico:2005dh}, the contributions to 
the Higgs mass--term of periodic and antiperiodic fields must be finely canceled 
in order to get a hierarchy between the compactification and EW scale. 
This adjustment was shown in \cite{Panico:2005dh} to be the main source of
fine--tuning. 
Thanks to the mirror symmetry, on the contrary, a natural partial cancellation
occurs in
the potential. We will see that a difference of an order of magnitude
between the compactification and EW scale, with $1/R\sim 1$ TeV, is
completely natural in our model. Such a scale is still too low to pass EWPT, but we significantly
lower the required amount of fine--tuning. 
Using the standard definition of \cite{Barbieri:1987fn}, thanks to the
mirror symmetry, we get a fine-tuning of $O(1\%)$, an
order of magnitude better than the $O(1\promille)$ found in \cite{Panico:2005dh}. 
If quantified with a more refined definition \cite{Anderson:1994dz} which takes into account
the presence of a possible generic sensitivity in the model, but under some physically motivated assumptions,
the fine-tuning turns out to be less, of about $O(10\%)$.
Another important difference with respect to the model of \cite{Panico:2005dh} is the choice
of the bulk fermion representations under the electroweak gauge group, which leads to a drastic reduction 
of the deviation to the SM $Z \overline b_L b_L$ coupling. This reduction is so effective, that the $Z \overline b_L b_L$
distortion becomes negligible in our set-up.

Interestingly enough, all SM states are even under the
mirror symmetry, so that the lightest odd particle is absolutely stable. In a large fraction of the
parameter space of the model, such a state is the first KK mode of an antiperiodic gauge field.
The mirror symmetry represents then an interesting way to get stable non--SM particles in non--universal extra dimensional
theories, where KK parity is not a suitable symmetry. Along the lines of \cite{Servant:2002aq}, it could be interesting
to investigate whether such state represents a viable Dark Matter (DM) candidate. 

Even assuming flavour and CP conserving new physics effects only, it
is necessary to add $18$ dimension $6$ operators to the SM in order 
to fit the current data (see {\it {e.g.}} \cite{Han:2004az,Grojean:2006nn}).
Among such 18 operators, it has recently pointed out that only 10 are sensibly constrained \cite{Cacciapaglia:2006pk}.
They can be parametrized by the seven universal parameters $\widehat S$, $\widehat T$, $\widehat U$,
$V$, $X$, $W$ and $Y$ introduced in \cite{Barbieri:2004qk}, the $\epsilon_b$ parameter \cite{Altarelli:1993sz},
and other two parameters which describe the deviation of the up and down quark couplings to the $Z$ boson.
Due to the way in which light SM fermions couple in our model and to the above suppression of the 
$Z \overline b_L b_L$ deviation,
it turns out that only the four parameters  $\widehat S$, $\widehat T$, $W$ and $Y$ are significant, like in a generic universal model as defined in \cite{Barbieri:2004qk}. 
Out of the several
parameters in our theory,  $\widehat S$, $\widehat T$, $W$ and $Y$ depend essentially only on $m_H$ and $1/R$.
The result of our fit is reported in fig.~3, which represents one of the main results of this paper.
We find that \mbox{$1/R\geq 4$ TeV} with a Higgs mass which can reach up to 600 GeV. 
Such high values of the Higgs mass are permitted because a relatively high
value of the $\widehat T$ parameter can be compensated by the effects
of a heavy Higgs (see \cite{Gogoladze:2006br} for a discussion of a
similar effect in the context of Universal Extra Dimensions (UED) \cite{Appelquist:2000nn}).
Although allowed, high values of $m_H$ are actually never reached in
our model, which predicts $100$ GeV $\leq m_H\leq 200$ GeV.
As can be seen from fig.~\ref{figEWPTfit}, our model can successfully
pass the combined fit of all electroweak precision observables.

The paper is organized as follows. In section 2 we introduce the model, emphasizing the main points in which
it differs from that of \cite{Panico:2005dh}. In section 3 we show our
results for the combined electroweak observables, including a brief discussion on the fine-tuning of the model. 
In section 4 we draw our conclusions and in the appendices
we collect some useful technical details.

\section{The Effective Lagrangian}

The 5D model we consider is a slight refinement of the one proposed in \cite{Panico:2005dh}.
The main essential feature is the introduction of a $\Z_2$ mirror symmetry.
The gauge group is taken to be
$G=SU(3)_w \times G_1 \times G_2$, where $G_i = U(1)_i \times SU(3)_{i,s}$ , $i=1,2$,
with the requirement that the Lagrangian is invariant under the $\Z_2$ symmetry $1\leftrightarrow 2$.
The periodicity and parities of the fields on the $S^1/\Z_2$ space
are the same as in \cite{Scrucca:2003ra,Panico:2005dh} for the electroweak $SU(3)_w$ sector, whereas
for the abelian $U(1)_i$ and non-abelian colored $SU(3)_{i,s}$ fields we now have (omitting for simplicity vector
and gauge group indices):
\be
A_1(y\pm 2\pi R) =  A_2(y) \,,
\;\;\;\;\;\;\;A_{1}(-y) =  \eta A_{2}(y) \,,
\label{bound-cond}
\ee
where $\eta_\mu=1$, $\eta_5=-1$, denoting by greek indices the 4D directions.
The unbroken gauge group at $y=0$ is $SU(2) \times U(1) \times G_+$,
whereas at $y=\pi R$ we have $SU(2) \times U(1) \times G_1\times G_2$,
where $G_+$ is the diagonal subgroup of $G_1$ and $G_2$. The $\Z_2$ mirror symmetry also survives the
compactification and remains as an exact symmetry of our construction.
It is clear from eq.~(\ref{bound-cond}) that the linear combinations $A_{\pm}=(A_1\pm A_2)/\sqrt{2}$
are respectively periodic and antiperiodic on $S^1$.
Under the mirror symmetry, $A_{\pm}\rightarrow \pm A_{\pm}$, so we can assign a
multiplicative charge +1 to $A_+$ and $-1$ to $A_-$.
The massless 4D fields are the gauge bosons 
in the adjoint of $SU(2)\times U(1)\subset SU(3)_w$, the $U(1)_+$ and gluon gauge fields $A_+$
and a charged scalar doublet Higgs field, arising from the internal
components of the odd $SU(3)_w$ $5D$ gauge fields, namely $A_{w}^{4,5,6,7}$.
The $SU(3)_{+,s}$ and $SU(2)$ gauge groups are identified respectively
with the SM $SU(3)_{QCD}$ and $SU(2)_L$ ones, while the hypercharge $U(1)_Y$ is
the diagonal subgroup of $U(1)$ and $U(1)_+$.
The extra $U(1)_X$ gauge symmetry which survives the orbifold
projection is anomalous (see \cite{Scrucca:2003ra,Panico:2005dh})
and its corresponding gauge boson gets a mass of the order of the cut-off scale $\Lambda$ of the
model.\footnote{In an interval approach, one can get rid of $U(1)_X$ by imposing Dirichlet
boundary conditions for it at one of the end-points of the segment. In fact, the two
descriptions are equivalent and become identical in the limit $\Lambda\rightarrow \infty$.}
An Higgs VEV, {\it i.e.} a VEV for the extra-dimensional components of
the $SU(3)_w$ gauge fields, induces the additional breaking to $U(1)_{EM}$.
Following \cite{Scrucca:2003ra,Panico:2005dh}, we parametrize this VEV
as
\be
\langle A_{w\,y}\rangle=\frac{2\alpha}{g_5 R}t^7\,,
\ee
where $g_5$ is the $5D$ charge of the $SU(3)_w$ group and $t^a$
its generators, normalized as $2\,\textrm{Tr}\,t^at^b=\delta^{ab}$ in the fundamental representation.
With this parametrization, the associated Wilson line is
\be
W=\displaystyle{e^{4 \pi i \alpha t^7}}\,.
\label{will}
\ee

We also introduce a certain number of couples of bulk fermions $(\Psi,\widetilde \Psi)$,
with identical quantum numbers and opposite orbifold parities. There are couples $(\Psi_1,\widetilde \Psi_1)$
which are charged under $G_1$ and neutral under $G_2$ and, by mirror symmetry, the same number
of couples $(\Psi_2,\widetilde \Psi_2)$ charged under $G_2$ and neutral under $G_1$.
No bulk field is simultaneously charged under both $G_1$ and $G_2$.
In total, we introduce one pair of couples $(\Psi_{1,2}^{t},\widetilde \Psi_{1,2}^{t})$ in the
anti-fundamental representation of $SU(3)_w$ and one pair of couples
$(\Psi_{1,2}^{b},\widetilde \Psi_{1,2}^{b})$ in the symmetric representation
of $SU(3)_w$. Both pairs have $U(1)_{1,2}$ charge +1/3 and are in the fundamental representation
of $SU(3)_{1,2,s}$.
The boundary conditions of these fermions follow from
eqs.~(\ref{bound-cond}) and the twist matrix introduced in \cite{Scrucca:2003ra}.
In particular, the combinations
$\Psi_{\pm}=(\Psi_1\pm\Psi_2)/\sqrt{2}$ are respectively periodic and antiperiodic on $S^1$.

Finally, we introduce massless chiral fermions with charge +1 with respect
to the mirror symmetry, localized at $y=0$.
As explained in \cite{Scrucca:2003ra,Panico:2005dh},
as far as electroweak symmetry breaking is concerned, we can focus on the
top and bottom quark only, neglecting all the other SM matter fields, which however can be accommodated
in our construction.
Mirror symmetry and the boundary conditions (\ref{bound-cond})
imply that the localized fields can couple only to $A_+$. Hence, we have an $SU(2)$
doublet $Q_L$ and two singlets $t_R$ and $b_R$, all in the fundamental representation of $SU(3)_{+,s}$
and with charge $+1/3$ with respect to the $U(1)_+$ gauge field $A_+$.

The most general 5D Lorentz breaking effective Lagrangian density, gauge invariant and mirror symmetric,
up to dimension $d<6$ operators, is the following (we use mostly minus
metric and $(\gamma^5)^2=1$):
\be
\mathcal{L}= \mathcal{L}_g+ \mathcal{L}_\Psi+\delta(y) \mathcal{L}_0+\delta(y-\pi R) \widehat{\mathcal{L}}_\pi\,,
\label{fullL}
\ee
with
\bea
\mathcal{L}_g= \a\a \sum_{i=1,2}\bigg[ -\frac{1}{2}{\rm Tr}\,G_{i\mu\nu}G^{i\mu\nu} -
\rho_s^2{\rm Tr}\,G_{i\mu5}G^{i\mu5} -\frac{1}{4}F_{i\mu\nu}
F^{i\mu\nu} -\frac{\rho^{2}}{2}F_{i\mu 5} F^{i\mu 5} \bigg] \nn \\
\a\a
-\frac{1}{2}{\rm Tr}\,F_{\mu\nu}F^{\mu\nu} -
\rho_w^2{\rm Tr}\,F_{\mu 5}F^{\mu5}\,, \label{Lgauge} \\
\mathcal{L}_\Psi = \a\a \sum_{i=1,2}
\sum_{a=t,b} \bigg\{\overline\Psi_i^{a} \big[i\Dslash_{4}(A_i)-k_a D_5(A_i)
\gamma^5 \big] \Psi_i^{a} \label{LPsi} \\
\a\a \ \ \ \
+\overline{\widetilde\Psi}_i^{a} \big[i\Dslash_{4}(A_i)-\widetilde k_a D_5(A_i) \gamma^5
\big] \widetilde\Psi_i^{a} -
M_a (\overline{\widetilde\Psi}_i^{a} \Psi_i^{a} +\overline{\Psi}_i^{a} \widetilde\Psi_i^{a})\bigg\} \,, \nn \\
\mathcal{L}_0= \a\a \overline Q_L i \Dslash_4(A_+) Q_L +
\overline t_R i \Dslash_4(A_+)t_R + \overline b_R i \Dslash_4(A_+) b_R
\nn \\
\a\a + \big(e_1^t \overline Q_L \Psi_{+}^{t} + e_1^b \overline Q_L \Psi_{+}^{b}
 + e_2^t \overline t_R \Psi_{+}^{t} + e_2^b \overline b_R \Psi_{+}^{b}+\mathrm{h.c.} \big)+
\widehat{\mathcal{L}}_0 \label{Lloc0}\,.
\label{LlocPi}
\eea
In eq.~(\ref{Lgauge}), we have denoted by $G_i=DA_{i,s}$ the gluon field strengths for $SU(3)_{i,s}$
and for simplicity we have not written the ghost Lagrangian and the gauge-fixing terms.
For the same reason, in eqs.~(\ref{LPsi}) and (\ref{LlocPi}) we have only
schematically written the dependencies of the covariant derivatives on the gauge fields
and in eq.~(\ref{Lloc0}) we have not distinguished the doublet and singlet
components of the bulk fermions, denoting all of them simply as $\Psi^{t}_+$ and
$\Psi^{b}_+$. Notice that $\Psi_+$ is the only bulk fermion that can have a mass-term mixing
 with the localized fields, since mixing with $\Psi_-$ and $\widetilde\Psi_-$ is forbidden by mirror symmetry
and the relevant components of $\widetilde\Psi_+$ vanish at $y=0$ due to the boundary conditions.
Extra brane operators, such as for instance localized kinetic terms,
are included in $\widehat{\mathcal{L}}_0$ and $\widehat{\mathcal{L}}_\pi$.
Additional Lorentz violating bulk operators like $\overline\Psi \gamma^5 \widetilde\Psi$,
$\overline\Psi \partial_y \Psi$ or $\overline\Psi i\Dslash_4 \gamma^5 \Psi$ can be forbidden by requiring
 invariance under the inversion of all spatial (including the compact one) coordinates,
under which any fermion transforms as $\Psi\rightarrow \gamma^0 \Psi$. 
This $\Z_2$ symmetry is a remnant of the broken $SO(4,1)/SO(3,1)$ Lorentz generators.
Notice that our choice of $U(1)$ charges allows mixing of the top quark with a bulk fermion in 
the ${\bf\ov 3}$ while the bottom couples with a ${\bf 6}$ of $SU(3)_w$. 
In \cite{Panico:2005dh}, the  choice of taking bulk fermions neutral under the $U(1)$ led to 
the opposite situation. As we will see, this greatly reduces
the deviation from the SM of the $Z \overline b_L b_L$ coupling, which
becomes negligible. 

Strictly speaking, the Lagrangian (\ref{fullL}) is not the most general one, since we are neglecting
 all bulk terms which are odd under the $y\rightarrow -y$ parity transformation and
can be introduced if multiplied by odd couplings. If not introduced, such couplings are not generated
and thus can consistently be ignored.

It is now possible to better appreciate the reason why we have introduced the above mirror symmetry.
In terms of periodic and antiperiodic fields,
the mirror symmetry constrains the Lorentz violating factors for periodic and antiperiodic fermions
to be the same: $k_+=k_-\equiv k$, $\widetilde k_+=\widetilde k_-\equiv \widetilde k$ for both the
${\bf \overline 3}$ and ${\bf 6}$ representations.
Finding an exact symmetry constraining these factors to be equal is important, since it was found
in \cite{Panico:2005dh} that the electroweak breaking scale was mostly sensitive to the ratio $k_+/k_-$,
which implied some fine-tuning in the model. Thanks to the mirror symmetry,
the fine-tuning is significantly lowered as we will better discuss in the subsection 3.1.
All SM fields are even under the mirror symmetry.
This implies that the lightest $\Z_2$ odd (antiperiodic)
state in the model is absolutely stable. This is a very interesting byproduct, since in a (large)
fraction of the parameter space of the model such state is the first KK mode of
the $A_-$ gauge field. The latter essentially corresponds to the first KK mode of the hypercharge
gauge boson in the context of UED \cite{Appelquist:2000nn},
which has been shown to be a viable DM candidate \cite{Servant:2002aq}.
Hence, it is not excluded that $A_-$ can explain the DM abundance
in our Universe. 

A detailed study of the model using the general Lagrangian (\ref{fullL}) is a too complicated task.
For this reason, we take $k_a=\widetilde k_a$ which considerably simplifies the analysis\footnote{It should be
emphasized that there is no fine-tuning associated to $k/\widetilde k$ (contrary to $k_+/k_-$)
and thus this choice represents only a technical simplification.} and set $\rho_w=1$.
The latter choice can always be performed without loss of generality by rescaling the compact
coordinate, and hence the radius of compactification as well as the other parameters of the theory.\footnote{Strictly speaking, in a UV completion
where the theory is coupled to gravity and the Lorentz violation is, for instance, due to
a flux background \cite{Panico:2005dh}, $\rho_w$ cannot be rescaled away
by redefining the radius of compactification, since
the latter becomes dynamical and $\rho_w$
essentially corresponds to a new coupling in the theory.
In our context, however, $R$ is simply a free parameter.}
Moreover, we neglect all the localized operators which are encoded in
$\widehat{\mathcal{L}}_0$ and $\widehat{\mathcal{L}}_\pi$.
The latter simplification requires a better justification that we postpone to the subsection 2.2.

\subsection{Mass Spectrum and Higgs Potential}

The localized chiral fermions, as explained at length in \cite{Scrucca:2003ra,Panico:2005dh}, introduce
gauge anomalies involving the $U(1)$ gauge field $A_X$. The effect
of such anomalies is the appearance of a large localized mass term for $A_X$ at $y=0$,
whose net effect is to fix to zero the value of the field at $y=0$: $A_X(0)=0$.
This complicates the computation of the gauge bosons mass spectrum, which is distorted
by this effect. The Lorentz violating factor $\rho^{2}$ also distorts the spectrum
of the KK gauge bosons, so that the analytic mass formulae for the $SU(3)_w\times U(1)_+$ gauge bosons are
slightly involved. They are encoded in the gauge contribution to the Higgs potential which we will
shortly consider. For $\rho =1$, the mass spectrum has been computed
in \cite{Panico:2005dh}.
The only unperturbed tower is the one associated to the $W$ boson,
whose masses are $m_W+n/R$ where
\be
m_W=\frac{\alpha}{R}\,,
\label{mw}
\ee
is the SM $W$ boson mass.
The mass spectra of the KK towers associated to the $U(1)_-$ gauge field $A_-$ and to the $\pm$ gluons
are trivial and given by $m_n=(n+1/2)\rho/R$ for $U(1)_-$, $m_n=n\rho_s/R$ for $SU(3)_{+,s}$ and
$m_n=(n+1/2)\rho_s/R$ for $SU(3)_{-,s}$.

The spectrum of the bulk-boundary fermion system defined by the Lagrangian (\ref{fullL})
is determined as in \cite{Panico:2005dh}, once one recalls the interchange between the fundamental
and symmetric representations of $SU(3)_w$, which simply lead to the
interchange $t\leftrightarrow b$ in eq.~(2.18) of \cite{Panico:2005dh}.
Let us introduce, as in \cite{Scrucca:2003ra,Panico:2005dh}, the
dimensionless quantities $\lambda^i = \pi R M_i$ and $\epsilon_i^a =
\sqrt{\pi R/2} e_i^a$. In the limit $\epsilon_1^t,\epsilon_2^t\gg 1$,
$\epsilon_1^b=\epsilon_2^b=0$, $\alpha\ll1$, one gets
\be
m_{t} \simeq  k_t m_W \frac{2 \lambda_t/k_t}{\sinh{(2\lambda_t/k_t)}}
\label{mtop-app}
\ee
which gives a top mass a $\sqrt{2}$ factor lighter than in \cite{Panico:2005dh}. We deduce
from eq.~(\ref{mtop-app}) that $k_t\sim 2\div 3$ is needed to get the top mass
in the correct range. 

For a large range of the microscopic parameters, the bulk-boundary fermion system also
gives the lightest new particles of our model. Such states
are colored fermions with a mass of order $M_b$, and, in particular, before Electroweak Symmetry Breaking (EWSB),
they are given by an $SU(2)$ triplet with hypercharge $Y=2/3$, a doublet
with $Y=-1/6$ and a singlet with $Y=-1/3$. For the typical values of the parameters
needed to get a realistic model, the mass of these states is of order $1-2\ \rm TeV$.

Let us now turn to the computation of the Higgs effective potential.
The fermion contribution in presence of bulk-to-boundary mixing and Lorentz violating
factors is the same as in \cite{Panico:2005dh}. The gauge contribution is slightly different,
because of the parameter $\rho^{2}$, which
in \cite{Panico:2005dh} was set to unity, for simplicity.
The one-loop gauge effective potential is however readily computed using the holographically inspired
method of \cite{Barbieri:2003pr} (see also \cite{Contino:2004vy} for a treatment of fermions in this context),
which also gives part of the mass spectrum. We refer the reader to appendix A for a brief review
of such technique, applied in our context.
The only gauge fields which contribute to the Higgs potential are $A_{w}$, associated to $SU(3)_w$,
and $A_+$, associated to $U(1)_+$. Before EWSB, it is trivial to integrate out the bulk at
 tree--level. In the mixed momentum-space basis, and in the unitary gauge, one finds the following holographic Lagrangian:
 \bea
 \label{holo1}
 {\mc L}_{\textrm {holo}} 
 \a = \a - \frac12 {\mc P}^{\mu\nu}
 \Big\{ {\textrm Tr}\left[
 -q \cot(2\pi R q)\widehat A_{w\,\mu}(-q)\widehat A_{w\,\nu}(q)+ q
 \csc(2\pi R q)\widehat A_{w\,\mu}(-q)P\widehat A_{w\,\nu}(q)P
 \right]  \nn \\
 \a+\a \left.\frac\rho2\left[ -q \cot(2\pi R q/\rho)\widehat
     A_{+\,\mu}(-q)\widehat A_{+\,\nu}(q)
 + q \csc(2\pi R q/\rho)\widehat A_{+\,\mu}(-q)\widehat A_{+\,\nu}(q)
 \right]
  \right\}
 \,,
 \eea
where $q=\sqrt{q_\mu q^\mu}$, ${\mc P}_{\mu\nu}=\eta_{\mu\nu} -q_\mu q_\nu/q^2$ is the standard transverse projector, $P=\textrm{diag}(-,-,+)$ is the orbifold projection matrix and $\widehat A$ are the holographic gauge fields,
as defined in appendix A. Since the Higgs is a Wilson line, its VEV
can be removed from the bulk through a non-single valued gauge
transformation \cite{hos}, which we
can choose in such a way that the boundary conditions at $y=\pi R$ remain
unchanged. In this new basis, 
we can obtain the Lagrangian after
EWSB directly from eq.~(\ref{holo1}) by replacing
 $A_{w}\rightarrow W^{1/2} A_{w}W^{-1/2}$, or $P\rightarrow PW$, where
 $W$ is the Wilson line as defined in eq.~(\ref{will}).
Notice that all the dependence of the action on $\alpha$, in the rotated field
  basis, comes from the boundary condition at $y=0$. When integrating
  out at tree--level the bulk and the $y=\pi R$
boundary, then, we are not neglecting any contribution to the one loop effective potential.
 After obtaining the holographic ($\alpha$--dependent) Lagrangian we have to impose the boundary
conditions, {\it i.e.} to put to zero the Dirichlet components of the holographic fields. We normalize the surviving (electroweak) gauge bosons in a non--canonical way (as in \cite{Barbieri:2004qk}) and then parametrize
\be
\widehat A_w  = \frac1{g_5}\left[
\sum_{a=1}^3 W^at_a+\frac1{\sqrt{3}} W^4t_8
\right]
\,,\;\;\;\;\;
\widehat A_+  =  \frac{\sqrt{2}}{\widetilde g_5}  W^4\,,
\label{holopar}
\ee
where $W^4$ indicates the hypercharge gauge boson and $\widetilde g_5$ 
the (common, due to mirror symmetry) $U(1)_{1,2}$ charges. 
We finally get an action of the form: 
\be
{\mc L}_{\textrm{holo}}=-\frac12 \sum_{i,j}W_\mu^i(-q)\Pi_{i,\,j}W^{\mu,j}(q)\,,
\label{holo2}
\ee
from which the gauge contribution to the one loop effective potential is found to be
\bea
V_g(\alpha)\a =\a -\frac32\int
\frac{d^4q}{(2\pi)^4}\log\left[\textrm{Det}\left(\Pi\right)\right]
= -\frac32 \int \frac{d^4q}{(2\pi)^4}\log\left[
\left(\cos{(2\pi\alpha)}-\cos{(2\pi q R)}\right)^{2}\right]\label{holo3}\\
\a -\a\frac32 \int \frac{d^4q}{(2\pi)^4}\log\left[
4 {g_5^\prime}^2\sin{(\pi q R)}\cos{(\pi q R/\rho)}\left(\cos{(4\pi\alpha)}-\cos{(2\pi q R)}\right)
\right.\nn\\
\a + \a 3{g_5}^2\rho\sin{(\pi q R/\rho)}\cos{(\pi q R)}\left(3+\cos{(4\pi\alpha)}-4\cos{(2\pi q R)} \right)
\Big]+\textrm{``$\alpha$-ind. terms"}\,.\nn
\eea
Note that the above integrals are divergent but they can be made
finite (after rotating to Euclidean momentum) by adding suitable $\alpha$--independent
terms which we have not written for simplicity. The zeroes of the
first square bracket in eq.~(\ref{holo3}) correspond to the
aforementioned $W$ tower while the second one provides the
mass-equation for the neutral gauge bosons sector. Notice that for 
$\rho\neq 1$ the $Z$ boson tower cannot be separated from the
ones associated to the photon and to $A_X$; all of them
arise from the zeroes of the second
term of eq.~(\ref{holo3}). The solution at $q=0$
corresponds to the physical massless photon while the first non-trivial
one determines the $Z$ boson mass.
We have checked that the
above mass equation can be also obtained, after a long calculation,
by directly computing the KK wave functions for the gauge bosons.

The full Higgs effective potential is dominated by the fermion contribution.
The presence of bulk antiperiodic fermions, whose coupling with the Higgs
are the same as for periodic fermions due to the mirror symmetry, allows for a natural
partial cancellation of the leading Higgs mass terms in the potential,
lowering  the position of its global minimum $\alpha_{min}$, as 
discussed in \cite{Panico:2005dh}, sec. 2.3.
\begin{figure}[t!]
\begin{minipage}[t]{0.465\linewidth} 
\begin{center}
\includegraphics*[width=\textwidth]{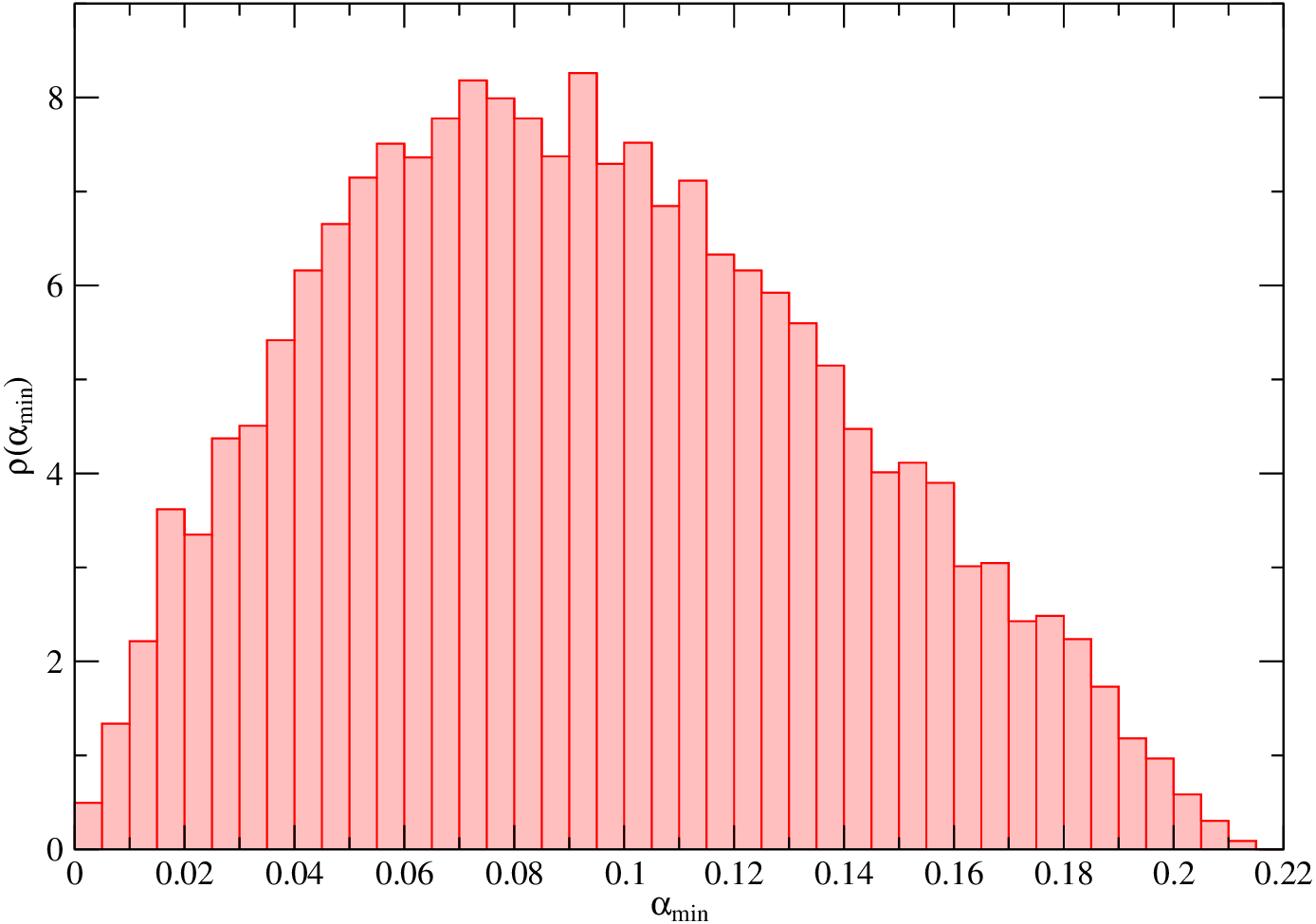}
\caption{Distribution of $\alpha_{min}$ for uniformely distributed input
parameters in the ranges $0.5<\rho<2$, $0.25<\lambda_{t,b}<1.25$,
$2<k_t<3$, $0.75<k_b<1.5$, $0.75<\epsilon_t^{1,2}<2.5$,
$0.1<\epsilon_b^{1,2}<0.45$.}\label{figHist}
\end{center}
\end{minipage}
\hspace{0.5cm} 
\begin{minipage}[t]{0.48\linewidth}
\begin{center}
\includegraphics*[width=\textwidth]{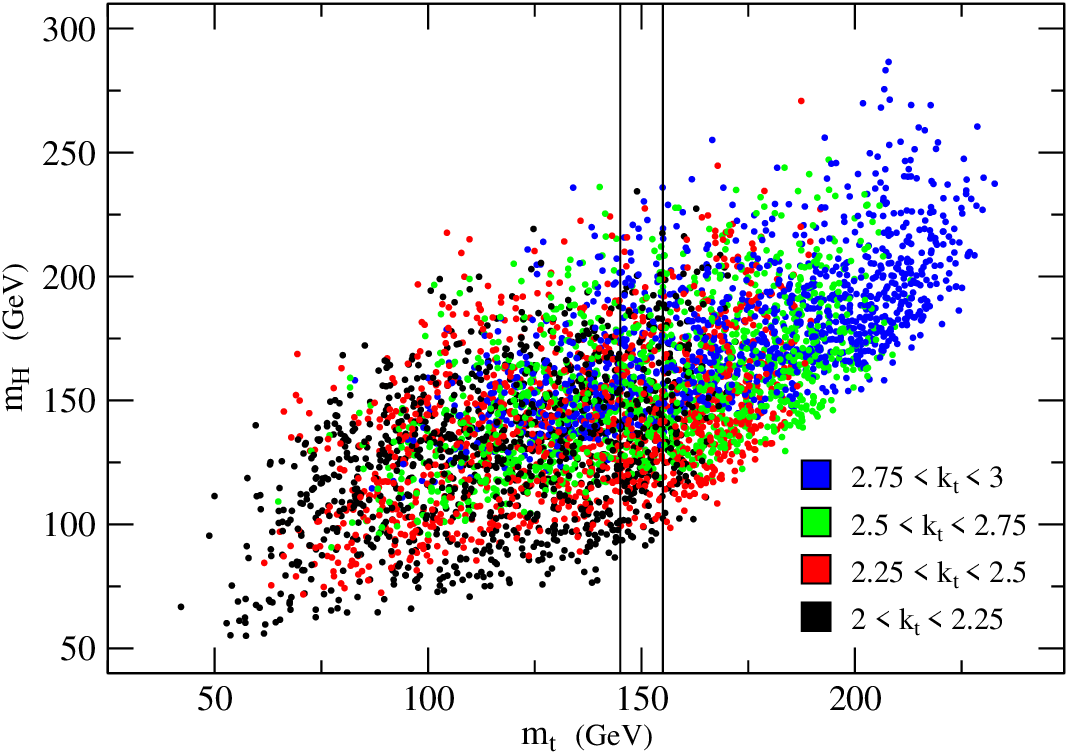}
\caption{Higgs and top masses for points with $\alpha_{min}<0.05$.
Different colors label different values of $k_t$. The
region among the two vertical black lines corresponds to the physical top mass.
}\label{figmhmtcolor}
\end{center}
\end{minipage}
\end{figure}
This can be seen from the histogram in fig.~\ref{figHist}, which shows the distribution
$\rho(\alpha_{min})$ for random
(uniformly distributed) values of the input parameters, chosen
in the ranges which give the correct order of magnitude for the top and bottom
masses.
We are neglecting the points with
unbroken EW symmetry ($\alpha_{min}=0$), which are about half of
the total.

{}From the effective potential we can determine the Higgs mass, which reads
\begin{equation}
m_H^2(\alpha_{min}) =
\displaystyle \frac{g_5^2 R}{8 \pi} \,
\frac{\partial^2 V}{\partial \alpha^2}\bigg|_{\alpha=\alpha_{min}} \;.
\label{MH}
\end{equation}
In fig.~\ref{figmhmtcolor} the Higgs mass is plotted versus the top
one and different colors label different ranges of the input parameter
$k_t$. We see how realistic values of $m_t$ can be obtained already
for $k_t=2$. The two vertical black lines in fig.~\ref{figmhmtcolor}
identify the region of physical top mass. This is taken to be around 150 GeV, which is the value
one gets by running the physical top mass using the SM RGE equations up to the scale $1/R$.
{}From fig.~\ref{figmhmtcolor} we see that the Higgs mass lies in the
$[100,200]$ GeV range. It is then significantly
lighter than what found in \cite{Panico:2005dh}.

\subsection{Estimate of the Cut-off and Loop Corrections}\label{secCutoffEstimate}

The model presented here, as any 5D gauge theory, is non-renormalizable and hence has a definite
energy range of validity, out of which the theory enters in a strong coupling uncontrolled regime.
It is fundamental to have an estimate of the maximum energy $\Lambda$ which we can probe
using our effective 5D Lagrangian (\ref{fullL}). First of all, we have to check that
$\Lambda$ is significantly above the compactification scale $1/R$, otherwise the 5D model
has no perturbative range of applicability. Once the order of magnitude of $\Lambda$ is
known, we can also estimate the natural values of the parameters entering in the Lagrangian
(\ref{fullL}). We define
$\Lambda$ as the energy scale at which the one-loop vacuum polarization corrections for the various gauge field
propagators is of order of the tree--level terms. This can be taken as
a signal of the beginning of a strong
coupling regime.
The relevant gauge fields are those of  $SU(3)_{i,s}$
and $SU(3)_{w}$, the abelian ones associated to $U(1)_{1,2}$ typically giving smaller corrections for
any reasonable choice of $\rho^{2}$.
Due to the Lorentz violation, however, the transverse ($A_\mu$)
and longitudinal ($A_5$) gauge bosons couple differently between themselves and with matter, so that
they should be treated separately.
We have computed the one-loop vacuum polarizations 
neglecting the effect of the compactification, which should give small finite size effects,
using a Pauli-Villars (PV) regularization in a non compact 5D space. We
denote by $\Lambda_s^{(\mu)}$,  $\Lambda_s^{(5)}$,
$\Lambda_w^{(\mu)}$ and $\Lambda_w^{(5)}$ the resulting cut-offs (due to the mirror symmetry,
the cut-offs associated to the two $SU(3)_s$ gauge groups coincide).
A reasonable approximation is to assume that the gauge contribution is approximately
of the same order as that of the fermions associated to the first two generations.
In this case,
by taking $k_t\simeq 2\div 3$, $k_b\simeq 1$, which are the typical phenomenological
values,
we get\footnote{Notice that the estimates (\ref{nda-cutoff}) differ
significantly from those reported in \cite{Scrucca:2003ra,Panico:2005dh}, since here
we have taken into account the fermion multiplicities. In addition,
the PV  regularization gives rise to a 5D loop factor in the vacuum polarization diagram which is $24\pi^2$ and
not $24\pi^3$, as taken in \cite{Scrucca:2003ra,Panico:2005dh} and naively expected.}
\bea
(\Lambda^{(\mu)}_s R)\frac{\alpha_s}{6}\frac 12 \Big(\frac{12}{k_t}+ \frac{24}{k_b}\Big)\sim 1 \ \ \Rightarrow
\ \ \Lambda^{(\mu)}_s \a\a \sim \frac{6}{R}\,, \nn \\
(\Lambda^{(5)}_s R)\frac{\alpha_s}{6}\frac 12 \Big(12k_t+ 24k_b\Big)\sim \rho_s^2 \ \ \Rightarrow
\ \ \Lambda^{(5)}_s \a\a \sim \frac{3\rho^2_s}{R}\,, \nn \\
(\Lambda^{(\mu)}_w R)\frac{\alpha_w}{6}\frac 12 \Big(\frac{12}{k_t}+ \frac{60}{k_b}\Big)\sim 1 \ \ \Rightarrow
\ \ \Lambda^{(\mu)}_w \a\a \sim \frac{5}{R}\,, \nn \\
(\Lambda^{(5)}_w R)\frac{\alpha_w}{6}\frac 12 \Big(12k_t+ 60 k_b\Big)\sim 1 \ \ \Rightarrow
\ \ \Lambda^{(5)}_w \a\a \sim \frac{4}{R}\,.
\label{nda-cutoff}
\eea
As can be seen from eq.~(\ref{nda-cutoff}), $\Lambda^{(\mu)}$ and
$\Lambda^{(5)}$ (both for the strong and weak coupling case) scale respectively as $k$ and
$1/k$ in the Lorentz violating parameters. This is easily understood by noting that the loop factor
scales as $1/k$ \cite{Panico:2005dh}. Thus $\Lambda^{(\mu)}\sim 1/\alpha \times k \sim k$ and
$\Lambda^{(5)}\sim 1/(\alpha\, k^2) \times k \sim 1/k$. Due to the group
factor of the $SU(3)_w$ symmetric representation, the electroweak
and strong interactions give comparable values for the cut-off.
Independently of the strong interactions,\footnote{Since $\rho_s^2$ is essentially a free parameter
in our considerations, we could anyhow increase $\Lambda^{(5)}_s$ by taking $\rho^2_s$ significantly larger than unity.}
the cut-off in the model is quite low: $\Lambda\sim 4/R$.
The doubling of the fields due to the mirror symmetry has the
strongest impact, since it decreases the electroweak cut-off estimate
by a factor of $2$. As far as the Lorentz violation is concerned, we see that eq.~(\ref{nda-cutoff}) gives, for
 $k_t=k_b=\rho_s=1$, $\Lambda_s^{(5)}\sim 3/R$.
 In the absence of Lorentz breaking and mirror symmetry, therefore, the final cut-off would have been
 as low as $3/R$. Even though eq.~(\ref{nda-cutoff}) only provides an order--of--magnitude estimate,
we can use it to place upper bounds on the allowed values of the Lorentz violating parameters
$k_t$ and $k_b$. To ensure $\Lambda\gtrsim 4/R$, we impose $k_t\lesssim3$ and $k_b\lesssim1.2$.

Once we have an estimate for the value of the cut-off in the theory, we can also
give an estimate of the natural size of the coefficients of the operators appearing
in the Lagrangian (\ref{fullL}). In particular, since we have
neglected them, it is important to see the effect of the localized
operators appearing in $\widehat{\cal{L}}_{0,\pi}$ when their
coefficients are set to their natural value.
At one--loop level, several localized kinetic operators are generated at the
fixed-points. In the fermion sector, which appears to be the most
relevant since it almost completely determines the Higgs potential, we
have considered operators of the form\footnote{Among all possible localized operators,
those with derivatives along the internal dimension require special care and are
more complicated to handle \cite{delAguila:2003bh}. It has been
pointed out in
\cite{delAguila:2006kj} that their effect can however be eliminated
by suitable field redefinitions (see also \cite{Lewandowski:2001qp}).}
\be
\widehat{\cal{L}}_{0,\pi} = \sum_i a^i_{0,\pi} \overline\psi^i i\,\dslash_4 \,
\psi^i\,,
\label{locop}
\ee
where $\psi^i$ indicates here the components of the bulk fields which are non-vanishing
at the $y=0$ or $y=\pi R$ boundaries. Such operators are
logarithmically divergent at one-loop level and thus not very sensitive to the cut-off scale;
anyhow, computing their coefficients $a^i_{0,\pi}$ using a PV regularization and setting $\Lambda \sim
4/R$, they turn out to be of order $10^{-3} R$.
By including these terms in the full Lagrangian, we have verified that the shape of the
effective potential, the Higgs and the fermion masses receive very
small (of order per-mille) corrections. Localized operators can however
sizeably affect the position of the minimum of
the effective potential when it is tuned to assume small
values. We will came back on this in subsection 3.1.

Finally, let us comment on the predictability of the Higgs mass at higher loops.
Although finite at one-loop level, the Higgs mass will of course develop divergencies
at higher orders, which require the introduction of various counterterms (see \cite{Maru:2006wa}
for a two-loop computation of the Higgs mass term).
The crucial point of identifying the Higgs with a Wilson line phase is the
impossibility of having a local mass counterterm at any order in perturbation theory.
This is true both in the Lorentz invariant and in the Lorentz breaking case.
For this reason, the Higgs mass is computable. In the Lorentz invariant case,
it weakly depends on the remaining counterterms needed to cancel divergencies loop by loop.
In the Lorentz breaking case, the Higgs mass still weakly depends on higher-loop counterterms,
with the exception of $\rho_w^2$, which is an arbitrary parameter directly entering in the Higgs mass formula. 
However, as we have already pointed out,
$\rho_w$ always enters in any physical observable together with $R$, so that it can be rescaled
by redefining the radius of compactification. Thanks to this property, the Higgs mass
remains computable also in a Lorentz non-invariant scenario.

\section{Phenomenological Bounds} 

A full and systematic analysis of all the new physical effects predicted by our
theory is a quite complicated task, mainly because it mostly depends on how flavour
is realized. For simplicity, we focus in the following on flavour and CP conserving new physics effects.
Due to the strong constraints on flavor changing neutral currents (FCNC)
and CP violation in the SM, this is a drastic but justified simplification,
since flavour does not play an important role in our mechanism
of EWSB. In fact we even did not discuss an explicit realization of it in our model
(see however \cite{Scrucca:2003ra,Martinelli:2005ix} for possible realizations and \cite{Panico:2005dh}
for an order of magnitude estimate of the bounds arising from FCNC). The only exception to this 
flavour universality is given by the third quark family, that has to be treated separately.

The analysis of flavour and CP conserving new physics effects requires in general the introduction of 18 dimension
6 operators in the SM to fit the data (see {\it {e.g.}} \cite{Han:2004az,Grojean:2006nn}).
It has recently been pointed out in \cite{Cacciapaglia:2006pk} that, out of these 18 operators,
only 10 are sensibly constrained. In a given basis (see \cite{Cacciapaglia:2006pk} for details),
$7$ of these operators are parametrized by the universal parameters $\widehat S$, $\widehat T$, $\widehat U$,
$V$, $X$, $W$ and $Y$ introduced in \cite{Barbieri:2004qk} extending the usual $S,T,U$ basis \cite{Peskin:1991sw}.
They are defined in general in \cite{Barbieri:2004qk}, starting from the inverse propagators $\Pi_{i,j}$
of the linear combinations of gauge bosons (not necessarily mass eigenstates of the Lagrangian) 
that couple to the SM fermions. As in the examples discussed in \cite{Barbieri:2004qk}, in our model 
these gauge fields coincide with the ``holographic'' electroweak bosons which appear in eq.~(\ref{holo2}). 
The remaining 3 operators are parametrized by the distortion $\delta
g_b$ (or the $\epsilon_b$ parameter \cite{Altarelli:1993sz}) of the $\widehat Z\, \overline b_L b_L$ coupling
and other two parameters which describe the deviation of the up and down quark couplings to the $\widehat Z$ boson.
As before, we put a hat superscript to distinguish the holographic field from its corresponding 5D field
or mass eigenstate. 
Due to the way in which SM fermions couple in our model, the later two parameters are totally negligible.
Being all light fermions almost completely localized at $y=0$ (see fig.~\ref{figlochist} in appendix B
for a quantitative idea of this effect), their couplings with the SM
gauge fields are universal and not significantly distorted.
As in other models based on extra dimensions \cite{Agashe:2003zs,Burd-Caccia,Agashe:2004rs},
a possible exception is the bottom quark coupling (the top even more, but its coupling to the $\widehat Z$ is
at the moment practically unconstrained), which is instead typically distorted by new physics.
In the original version of our model considered in \cite{Panico:2005dh}, indeed, $\delta g_b$
turned out to give one of the strongest constraints on the model.
As anticipated before and as we will now see in some detail, the simple idea of reversing
the mixing of the localized bottom and top quark with respect to \cite{Panico:2005dh}
strongly reduces $\delta g_b$.\footnote{We would like to thank G. Cacciapaglia for this observation.}

To be more precise, the distortion of the $\widehat Z\, \overline b_L b_L$ coupling is due to two
effects.
One of those is the mass--term mixing of the $b_L$ with the KK tower of the $SU(2)_w$
triplet and singlet fermions coming from the bulk field in the rep. $\bf 6$ of
$SU(3)_w$. The second distortion is a consequence of the delocalization of $b_L$, which then
also couples to gauge bosons components orthogonal to $\widehat Z$. Both effects are proportional to the
mixing parameters $\epsilon_1^{t,b}$.
Interestingly enough, at leading order in $\alpha=m_W R$ ,
the former distortion exactly vanishes in the (very good) approximation of neglecting
the bottom mass. This can be understood by noting
that the $b_L$ mixes with the component of an $SU(2)_w$ triplet which
has $T_3=-1$ and with a singlet ($T_3=0$). The mixing with the triplet state leads to an increase
(in magnitude) of $g_b$, whereas the singlet makes it decrease. The two effects turn out to
exactly compensate each other.
The distortion due to the delocalization of $b_L$
is instead non-vanishing. An explicit computation along the lines of the one reported in appendix A gives, 
at leading order in $\alpha$:
\be
\delta g_b = g_b -g_b^{SM} =\frac{g_b^{SM}\pi^2 \alpha^2}{4 Z_1^b \cos^2\theta_W}\sum_{i=t,b}
\frac{{\epsilon_1^i}^2}{\lambda_i^2} F\left(\frac{\lambda_i}{k_i}\right)\,,
\label{deltaZbb}
\ee
where $F(x)$ is defined in eq.~(\ref{Fx-Zbb})
and $Z_{1}^b$ is the wave function normalization factor needed to canonically normalize the $b_L$ field,
defined in the equation (2.18) of \cite{Panico:2005dh} (recall $t\leftrightarrow b$
with respect to \cite{Panico:2005dh}).
We have verified the validity of eq.~(\ref{deltaZbb})
by computing it both holographically and from a more direct KK analysis. In the latter case, 
the distortion is caused by the KK tower of the $A_X$ gauge field, which mixes to the
lowest mass eigenstate of the $Z$ boson and couples to $b_L$ in the bulk. 
It is worth to notice that the two results agree completely, but only when one sets the $\widehat Z$ boson on-shell
in the holographic approach, as it should be done. This result is expected, if one considers that at the pole, the
$\widehat Z$ gauge boson is totally given by its lowest mass eigenstate $Z$.\footnote{Strictly speaking, due to the $Z$ width,
there will still be left a combination of the other KK eigenstates, but this effect, not visible at tree--level, is totally negligible.}
In the holographic approach, as shown in appendix A in a toy example,
one would find $\delta g_b$ to be identically zero if it is computed at zero momentum.
Since $F(x)$ is a positive function, $\delta g_b<0$.
In the typical range of parameters of phenomenological interest, $-\delta g_b \lesssim 2 \alpha^2$
and it comfortably lies inside the experimentally allowed region (at $1\ \sigma$
form the central value) if $\alpha \lesssim 3\times10^{-2}$.
As we will now discuss, stronger bounds on $\alpha$ arise from universal corrections and
non-universal corrections to $Z\ov{b}_L b_L$ can be neglected in our global fit.

We are left with the 7 parameters $\widehat S$, $\widehat T$, $\widehat U$,
$V$, $X$, $W$ and $Y$. From the inverse gauge propagators $\Pi_{i,j}$
in eq.~(\ref{holo2}), one easily finds, at tree--level and at leading order in $\alpha$, 
\be
\left\{
\begin{array}{l}
\widehat S =\displaystyle\frac23\alpha^2\pi^2 \\
\rule{0pt}{1.5em}\widehat T = \alpha^2\pi^2\\
\rule{0pt}{1.5em}\widehat U = \displaystyle-\frac23 \alpha^4\pi^4\\
\rule{0pt}{1.5em}V = \displaystyle\frac{14}{45}\alpha^6\pi^6
\end{array}
\right.\,,
\hspace{3.5em}
\left\{
\begin{array}{l}
X = \displaystyle\frac{14}{45}\tan(\theta_w)\alpha^4\pi^4\\
\rule{0pt}{2.25em}Y =\displaystyle\frac{\rho^2\sin^2(\theta_w)+1+2\cos(2\theta_w)}{
9\rho^2\cos^2(\theta_w)}\alpha^2\pi^2 \\
\rule{0pt}{1.75em}W = \displaystyle\frac13\alpha^2\pi^2
\end{array}
\right.\,.
\label{obliquepar}
\ee
Since $\alpha$ has to roughly be of order $10^{-2}$, we see from eq.~(\ref{obliquepar})
that the 3 parameters  $\widehat U$, $V$ and $X$ are totally negligible. This is actually
expected for any universal theory \cite{Barbieri:2004qk}.
Notice that the universal parameters essentially depend only on $\alpha$, the other parameter 
entering is $\rho$, which affects only $Y$.
We have verified that radiatively generated localized gauge kinetic terms give negligible effects
in eq.~(\ref{obliquepar}) and can thus be consistently neglected.

\begin{figure}[t!]
\begin{center}
\includegraphics*[width=.65\textwidth]{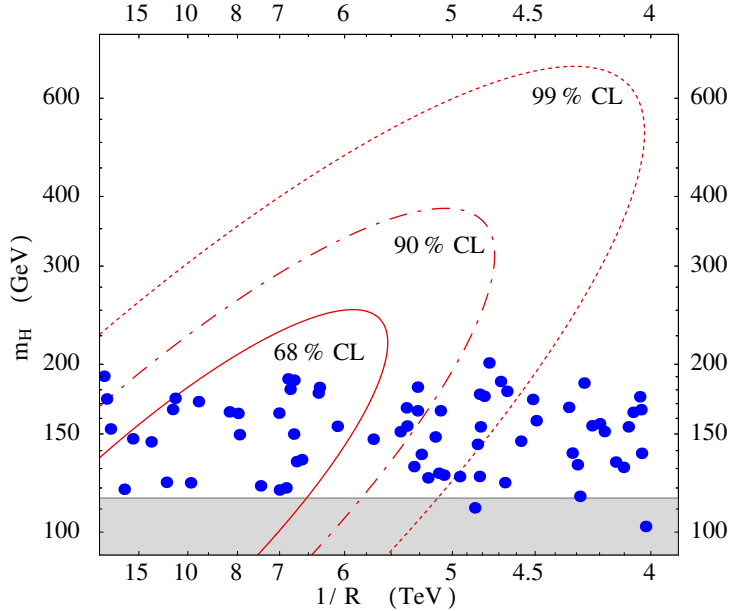}
\caption{Constraints coming from a $\chi^2$ fit on the EWPT. The contours represent
the allowed regions in the $(1/R, m_H)$ plane at $68\%$, $90\%$ and $99\%$
confidence level (2 d.o.f.). The shaded band shows the experimentally
excluded values for the Higgs mass ($m_H < 115\ \rm GeV$). The blue dots
represent the predictions of our model for different values of the microscopic
parameters (only points with the correct top and bottom masses
are plotted).}\label{figEWPTfit}
\end{center}
\end{figure}
In fig.~\ref{figEWPTfit}, we report the constraints on the Higgs mass and the
compactification scale due to all electroweak flavour and CP conserving observables,
obtained by a $\chi^2$ fit using the values in eq.~(\ref{obliquepar}).\footnote{We would like
to thank A. Strumia for giving us the latest updated results for the $\chi^2$ function obtained
from a combination of experimental data (see table 2 of \cite{Barbieri:2004qk} for a detailed list of
the observables included in the fit), as a function of the universal parameters of eq.~(\ref{obliquepar}).}
To better visualize the results, we fixed the value of the $\rho$
parameter to the Lorentz-symmetric value ($\rho=1$), hence we used a $\chi^2$ fit with 2
d.o.f. The fit essentially does not depend on $\rho$, as can be checked determining $\rho$ by a minimization of
the $\chi^2$ function, which gives a plot which is almost indistinguishable from that reported in fig.~\ref{figEWPTfit}.
Due to their controversial interpretation (see {\it {e.g.}} \cite{Altarelli:2004fq}), we decided to exclude from the fit
the NuTeV data. Nevertheless, we verified that the inclusion of such experiment leaves our results
essentially unchanged. From fig.~\ref{figEWPTfit} one can extract a lower bound on the
compactification scale $1/R \gtrsim 4-5\ \rm TeV$ (which corresponds to
$\alpha \lesssim 0.016-0.02$) and an upper bound on the Higgs mass which varies from
$m_H \lesssim 600\ \rm GeV$ at $1/R \sim 4\ \rm TeV$ to $m_H \lesssim 250\ \rm GeV$ for
$1/R \gtrsim 10\ \rm TeV$. 
Notice that the values for $\widehat S$, $W$ and $Y$ (for $\rho=1$) found in eq.~(\ref{obliquepar}) are exactly those expected in a generic $5D$ theory with gauge bosons and Higgs in the bulk, as reported in eq.~(20) of 
\cite{Barbieri:2004qk}. On the contrary, the $\widehat T$ parameter in eq.~(\ref{obliquepar}) is much bigger and is essentially due to the anomalous gauge field $A_X$. A heavy Higgs is allowed in our model because of this relatively high value of the $\widehat T$ parameter, which compensates the effects of a high Higgs mass.

Analyzing the models obtained by a random scan of the microscopic parameters, we found that
the bound on the compactification scale can be easily fulfilled if a certain
fine-tuning on the parameters is allowed. On the other hand, the bounds on the Higgs mass
does not require any tuning. 

\subsection{Sensitivity, Predictivity and Fine-Tuning}

As expected, the constraints arising from the electroweak observables require a quite stringent
bound,  $1/R\gtrsim 4.5\ \rm TeV$, on the size of the compactification
scale. Differently from previously considered scenarios of GHU in flat space (see {\it e.g.} \cite{Scrucca:2003ra}), our
model can satisfy such a bound (see fig.~\ref{figEWPTfit}) for
suitable choices of the microscopic input parameters.
It is clear, however, that a certain cancellation (fine-tuning) must
be at work in the Higgs potential to make its minimum $\alpha=m_W R$ 
as low as $0.018$.

\begin{figure}[t]
\begin{minipage}[t]{0.48\linewidth}
\begin{center}
\includegraphics*[width=\textwidth]{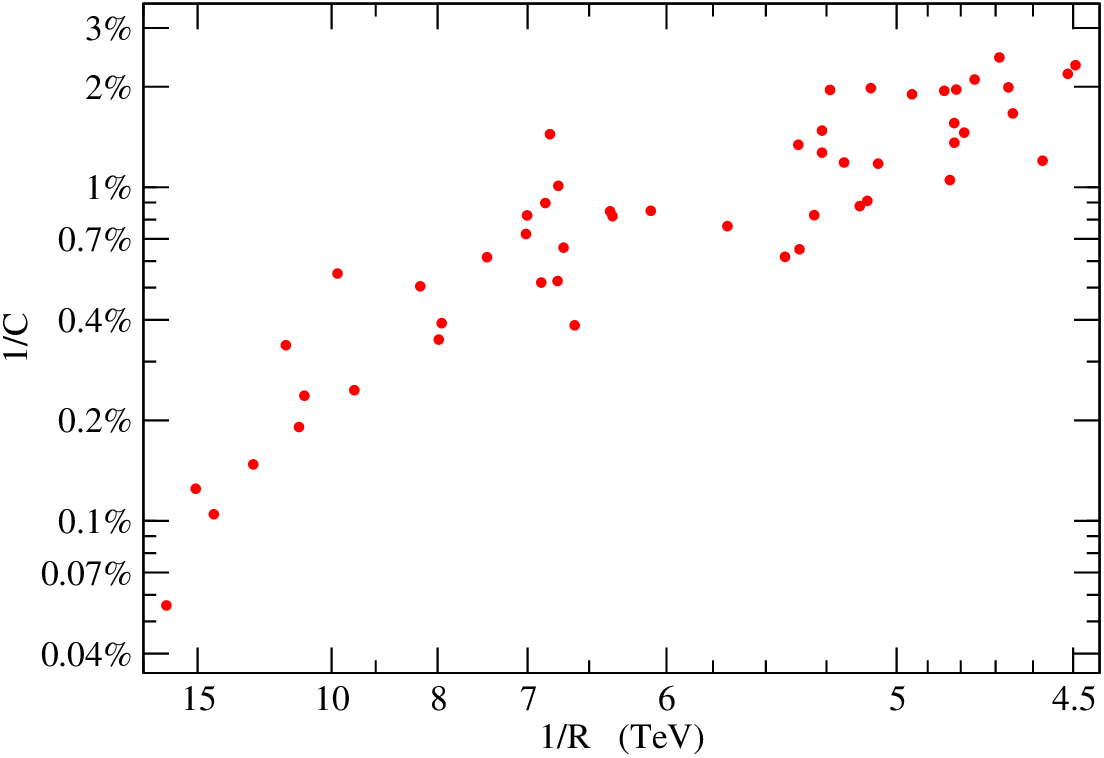}
\caption{Inverse sensitivity for those points which pass the $\chi^2$ test.}\label{figInverseC}
\end{center}
\end{minipage}
\hspace{0.5cm}
\begin{minipage}[t]{0.47\linewidth}
\begin{center}
\includegraphics*[width=\textwidth]{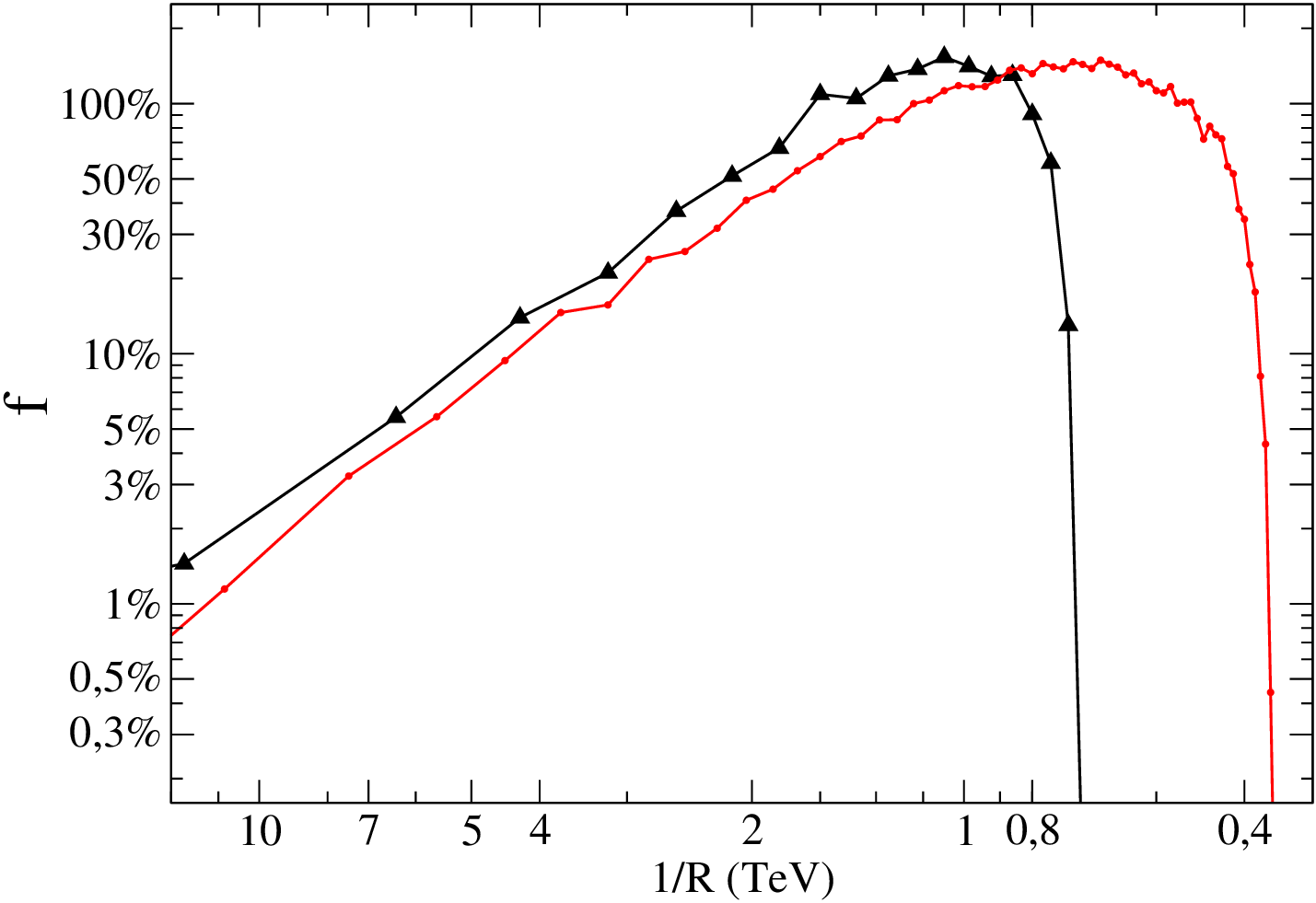}
\caption{Bayesian measure of the fine-tuning
at fixed (black) and not fixed (red) top mass.}
\label{figcastano}
\end{center}
\end{minipage}
\end{figure}

The fine-tuning is commonly related \cite{Barbieri:1987fn} to the
sensitivity of the observable ($\alpha$, in our case) with respect to
variations of the microscopic input parameters.\footnote{
In the particular case at hand, the sensitivity of $\alpha$ is related 
to the cancellation of the Higgs mass term in the effective potential. As a consequence, one 
expects the amount of tuning to scale roughly as $\alpha^2$, as confirmed numerically.
} We define
\be
C \equiv {\textrm{Max}}\left\{\left| \frac{\partial \log
      \alpha}{\partial \log I}\right|\right\}
\,,
\ee
as the maximum of the logarithmic derivatives of $\alpha$ with
respect to the various input parameters $I$. The fine-tuning, defined 
as $f=1/C$, is plotted in fig.~\ref{figInverseC} for the  
points which pass the $\chi^2$ test.
For $1/R\sim 4.5\ \rm TeV$, the fine-tuning of our model is $1\,\%$.

According to a more refined definition \cite{Anderson:1994dz}, the
fine-tuning should be intended as an 
estimate of how unlike is a given value for an observable. 
This definition is not directly
related to sensitivity, even though it reduces to $f=1/C$ in many
common cases. In this framework, the fine-tuning is computed from the
probability distribution $\rho(\alpha)$ of the output observable
for ``reasonable'' distributions of the input microscopic
parameters. One defines \cite{Anderson:1994dz}
\be
f=\frac{\alpha\rho(\alpha)}{\langle\alpha\rho(\alpha)\rangle}\,,
\label{fcast}
\ee
where\footnote{
Our definition of $\langle\alpha\rho\rangle$ is not precisely 
the one given in \cite{Anderson:1994dz}. The difference however
disappears when, as in our case, the input variables are uniformely distributed.
} $\langle\alpha\rho(\alpha)\rangle=\int d\alpha\,\alpha\left[\rho(\alpha)\right]^2$.
The choice of the range of variation of the input parameters\footnote{
We have checked that our results are independent on the detailed form
of the probability distribution of the input. This is because the ranges which we
consider are not so wide. Uniform distributions have been used to derive
the results which follow.} 
is the strongest ambiguity of the procedure, and must reflect
physically well motivated assumptions. 
The distribution shown in fig.~\ref{figHist},
which we will use, has been obtained with quite generic ranges, which however
reflect our prejudice on the physical size of the top and bottom quark
masses. Even though this does not make an important difference 
(see fig.~\ref{figcastano}), we can make our assumption more precise by
further restricting to points in which the top (not the b, due to the
lack of statistic) has the correct mass. 
The straightforward application of eq.~(\ref{fcast}) gives rise to
the fine-tuning plot of fig.~\ref{figcastano}.
At $1/R= 4.5\ \rm TeV$ this prescription gives a fine-tuning of
$10\%$. One could also relax the assumption of having EWSB.
Since the fraction of  points without EWSB is about one half of the total,
one could argue, very roughly, that the resulting fine-tuning will increase of a factor two
and be of order 5$\%$. This result strongly depends on the
assumption we did on the fermion mass spectrum.
In practice, what we observe is a certain correlation among the
requirements of having a massive top and an high
compactification scale, in the sense that points with massive top (and
light b) also prefer to have small minima.

Independently on how we define the fine-tuning, however, the high value of $C$
at small $\alpha$ represents a problem by itself.
The high sensitivity, indeed, makes $\alpha$ unstable
against quantum corrections or deformations of the Lagrangian with the
inclusion of new (small) operators.
As described in section~\ref{secCutoffEstimate}, we have considered the
effects on the observables of localized kinetic terms for bulk
fermions. We found that such
operators, which lead to very small corrections to all other observables
considered in this paper, can completely destabilize the compactification
scale at small $\alpha$. 
A similar effect is found when including in the effective
potential the very small contributions which come from the light
fermion families. For $\alpha\sim0.02$ the corrections which
come in both cases are of
order $50\%$, so that the compactification radius is effectively not
predicted in terms of the microscopic parameters of the model.

\section{Conclusions}

In this paper we presented a realistic GHU model on a flat 5-dimensional
orbifold $S^1/\Z_2$. The model is based on the one outlined in~\cite{Panico:2005dh}, in which
an explicit breaking of the $SO(4,1)$ Lorentz symmetry down to the usual 4-dimensional
Lorentz group $SO(3,1)$ was advocated in order to overcome some of the most worrisome
common problems of GHU models on flat space. 

The main new feature of the present model is the presence of a $\Z_2$ ``mirror symmetry'',
which essentially consists in a doubling of part of the bulk fields, 
contributing to a
reduction of the fine-tuning. It also allows to have a stable boson at the $\rm TeV$ scale, 
which could prove to be a viable dark matter candidate.
We have also shown how, due to a suitable choice of the bulk fermion
quantum numbers,
the deviations from the $Z \overline{b}_L b_L$ SM coupling can be made totally negligible.
The new physics effects can be encoded in the 4 parameters $\widehat S$,
$\widehat T$, $W$ and $Y$ defined in~\cite{Barbieri:2004qk}, and thus our model
effectively belongs to the class of universal theories. From a combined
fit, we derived a lower bound on the compactification scale
$1/R \gtrsim 4\ \rm TeV$ with allowed (although never reached, see fig.~3)
Higgs masses up to 600 GeV.
Our model is compatible with the experimental constraints if a certain tuning in the parameter space, 
of order of a few $\%$, is allowed. Such a fine-tuning is certainly acceptable, but it is nevertheless
still too high to claim that our model represents a solution to the little hierarchy problem.
However, it should be emphasized that the fine-tuning we get is of the same order of that
found in the MSSM. 

\section*{Acknowledgments}

We would like to thank G. Cacciapaglia for participation at the early stages of this work and
A. Strumia for providing us the complete $\chi^2$ function used in this work and for useful discussions. 
We would also like to thank C. Grojean, A. Pomarol, R. Rattazzi and A. Romanino for interesting discussions.
This work is partially supported by the European Community's Human
Potential Programme under contract MRTN-CT-2004-005104 and by the Italian
MIUR under contract PRIN-2003023852.

\appendix

\section{The ``Holographic'' Approach }

Deriving $4D$ effective theories from $5D$ models is usually done by
integrating out the massive KK eigenstates, keeping the lightest states
of the KK towers as effective degrees of freedom.
When the extra space has a boundary at $y=0$, an alternative and useful
 ``holographic'' possibility is to use the boundary values 
$\widehat\Phi(x)=\Phi(x,0)$ of the $5D$ fields 
as effective
degrees of freedom and to integrate out all the fields components
$\Phi(x,y)$ with $y\neq 0$ \cite{Barbieri:2003pr}. As long as
$\Phi(x,0)$ has a non-vanishing component of the light mode of the KK 
tower associated to $\Phi(x,y)$, the holographic and KK approaches are
completely equivalent.
In warped 5D models,
such a prescription closely resembles the one defined by the AdS/CFT correspondence \cite{Ads-cft}, 
where roughly speaking $\widehat\Phi(x)$ is the source of a CFT operator and $\Phi(x,y)$ is the corresponding
AdS bulk field. Although in flat space such interpretation is missing,
it has been emphasized in \cite{Barbieri:2003pr}
that it is nevertheless a useful technical tool to analyze various properties of 5D theories.
The holographic approach, in this paper, has been used to perform the
global fit to EW observables. Along the lines of
\cite{Barbieri:2004qk}, it significantly simplifies the analysis.

We think it might be useful to collect here a few technical details on
how to use this approach in our particular context. To keep the discussion as clear
as possible, and to emphasize the key points, we consider in this
appendix a simple $5D$ Lorentz invariant
model on a segment of length $L$, which resembles some aspects of our construction. 
Let $A_M$ be an $U(1)$ bulk gauge field, $\psi$ and $\chi$ a couple of
bulk charged fermions and $q_R$ a chiral $4D$ fermion localized at $y=0$,
which mixes with $\psi_L$. The gauge field components $A_\mu$ and
$A_5$ satisfy respectively 
Neumann amd Dirichlet
boundary conditions at $y=L$, whereas for the fermions we have
$\psi_R(L)=\chi_L(L)=0$. 
For what concerns the bosonic fields, 
boundary conditions at $y=0$
are simply imposed
by the choice of the dynamical holographic fields to be retained. 
The holographic approach to fermions is a bit
more complicated since one has to modify the action by introducing
suitable localized mass-terms \cite{Contino:2004vy}. With the convention 
$\gamma^5\psi_L=+\psi_L$, the action reads
\bea
S= \int_0^L \! dy\a\a\!d^4x \bigg\{-\frac 14 F_{MN}F^{MN}+
\Big[\frac12 (\overline\psi i\Dslash_5 \psi + \overline{\chi} i\Dslash_5 \chi)
- M \overline{\chi} \psi + h.c. \Big]\bigg\}
 \nn \\
+\int\!d^4x\a\a\frac12 \Big[\overline q_R i\Dslash_4 q_R + e\, \overline q_R \psi_L(0) +
e\, \overline \psi_L(0) q_R
+\overline\psi(0)\psi(0)-\overline\chi(0)\chi(0)
\Big] \,,
\label{toyaction}
\eea
and one can check that, by performing generic variation of the bulk fermion
fields, the boundary
conditions $\psi_R(0)=-e/2\,q_R$ and $\chi_L(0)=0$ automatically
arise as equations of motion.
The field which we will retain are
$\widehat{A}_\mu=A_\mu(0)$ and $q_R$.
Integrating out at tree-level the 
remaining degrees of freedom is equivalent
to solve their equation of motion with fixed holographic
fields
($\delta\widehat{A}_\mu=\delta q_R=0$
in the action variation)
 and to plug the solution back into the action. Varying
 eq.~(\ref{toyaction}) with the holographic prescription simply gives
 the standard $5D$ Dirac and Maxwell equations plus the above
 mentioned boundary conditions for the fermions.

 It is useful to adopt a mixed
momentum-space basis $\phi(p,y)$ for the fields.
In the unitary gauge $A_5$ vanishes and the longitudinal component of $A_\mu$
is decoupled, while the equations for the transverse part $ A_{\mu}^t =
{\cal P}_{\mu\nu} A^\nu$ (${\cal P}_{\mu\nu}= \eta_{\mu\nu} - p_\mu
p_\nu/p^2$) 
reduce to $(p^2+\partial_y^2) A_{\mu}^t(p,y)=0$.
Imposing the boundary condition $\partial_y A_\mu(L)=0$, one easily gets
\be
A_\mu(p,y) =
\frac{\cos\left[p(L-y) \right]}{\cos\left[p\,L\right]}\widehat{A}_{\mu}^t(p)\,,
\label{EOM-A}
\ee
where $p=\sqrt{p^\mu p_\mu}$. One proceeds analogously for fermions. 
It is a simple exercise, given the boundary conditions at $y=L$, to
determine $\psi(p,y)$ 
and $\chi(p,y)$  in terms of $q_R$. One gets
\bea
\psi_R \a = \a -\frac{e}{2}\, \frac{\sin\left[\omega(L-y)
  \right]}{\sin\left[\omega\,L\right]} 
\, q_R\,; \; \; \; \; \; 
\psi_L =-\frac{e}{2}\, \frac{\cos\left[\omega(L-y)
  \right]}{\omega \sin\left[\omega\,L\right]}\, \pslash\, q_R \,; \nn \\
\chi_L \a = \a 0\,; \;\;\;\;\;\;\;\; \;\;\;\;\; \;\;\;\;\;\;\;\;\;\;\; \;\;\;\;\;\;\;\;\;\;\;
\chi_R =-\frac{e}{2}\, \frac{\cos\left[\omega(L-y)
  \right]}{\omega\sin\left[\omega\,L\right]}\,M\, q_R\,; 
\label{EOM-Psi}
\eea
where $\omega = \sqrt{p^2-M^2}$. Plugging back the solutions
(\ref{EOM-A}) and 
(\ref{EOM-Psi}) into the
action (\ref{toyaction}), we can get the holographic
Lagrangian. At quadratic level
\be
{\cal L}_{holo} = -\frac 12 \Pi(p) \widehat{A}_{\mu}^t
\widehat{A}^{t\,\mu}
+ \frac12 Z_q \overline q_R \,\pslash_4 \, q_R\,,
\label{Loloquad}
\ee
where we defined
\begin{equation}
\Pi(p) = p \tan p L, \ \ \ Z_q =  1-\frac{e^2}{2}\, \frac{\cot \omega L}{\omega} \,.
\label{WF-terms}
\end{equation}
 
The holographic procedure can also be used to study interaction terms. Again, one has simply
to substitute into the action (\ref{toyaction}) the solutions
(\ref{EOM-A}) and (\ref{EOM-Psi}). 
It is useful to see how this works by computing the
distortion to the $U(1)$ gauge coupling $g$ of $q_R$ to $\widehat A_\mu$ due to the massive modes.
This interaction is encoded in the following Lagrangian term:
\bea
{\cal L}^{(3)} = \a\a g \int^L_0\!dy \Big[\overline \psi (p+q,y)  \Aslash(q,y) \psi(p,y)+
\overline{\chi} (p+q,y)  \Aslash(q,y) \chi(p,y) \Big] \nn \\ 
\a\a +\, \frac{g}{2}\, \overline q_R(p+q) \widehat \Aslash(q) q_R(p) \,.
\label{3-vertex}
\eea
In order to simplify our computation, let us consider the kinematic
configuration in which the fermion is on-shell, $p^2=(p+q)^2=0$, and 
$q^2=\mu^2\ll M^2$. By direct substitution and after a bit of algebra, 
one gets the following cubic interaction, up to terms of order $\mu^2/M^2$:
\bea
{\cal L}^{(3)} = \a\a  \frac{g}{2}
\Big[Z_q + \frac{e^2\,\mu^2}{8 M^2} F(ML)\Big] \overline q_R(p+q) \widehat \Aslash^t(q) 
q_R(p) \,,
\label{3-vertex-corr}
\eea
where $Z_q$ is computed at zero momentum and we have defined
\begin{equation}
F(x) \equiv 1 - \frac{1}{x}\coth x + \coth^2 x\,.
\label{Fx-Zbb}
\end{equation}
As expected, the form of the interaction vertex at $\mu^2=0$ has
precisely the same structure as 
the quadratic term
in (\ref{Loloquad}), as required by gauge invariance, which forbids any correction to $g$. 
At quadratic order in the gauge boson momentum, however, we get a correction
\begin{equation}
\delta g = \frac{g\,e^2}{8Z_q} \frac{\mu^2}{M^2} F(ML)\,.
\label{deltagU(1)}
\end{equation}
Notice that eq.~(\ref{deltagU(1)}) coincides exactly with eq.~(\ref{deltaZbb}), when fixing $\mu^2=m_Z^2$ and taking into account
the SM coupling of the $b_L$. This is a further proof that no
corrections arise in our model from the mixing of $b_L$
with fields with different isospin quantum numbers, the only effect being given by the
partial delocalization of the field, resulting in couplings with the ``massive'' gauge fields $A_\mu(y)$, with
$y\neq0$. 

The generalization to the non-abelian case and fields in different representations of the gauge group does not present
any conceptual problem. When a Wilson line symmetry breaking occurs, like in our model, the most convenient
thing to do is to go in a gauge in which the VEV for $A_5$ vanishes and the boundary conditions for the various fields
are twisted at $y=0$. In this way, the effect of the twist simply amounts to a redefinition of the holographic fields,
as discussed in section 2.1.

\section{Fermion Wave Functions}

In this appendix we turn to the more standard analysis of the KK mass
eigenstates and examine the wave functions of the fermions arising from the
mixing of the bulk fields with the fields localized at $y=0$. 
The purpose of this analysis is to illustrate how much the SM fermions are localized in our model.
As we will see, in fact, the top quark is not localized at all, whereas the bottom
is only partially localized (see figs.~\ref{figfdotb} and \ref{figlochist}).
In order to simplify the discussion, we focus on the wave functions of right-handed singlets before EWSB.
Left-handed localized fields are more involved since they couple to two different bulk fields,
as shown in eq.~(\ref{Lloc0}).

The relevant quadratic Lagrangian describing the coupling of a localized right-handed fermion
with the bulk fields is easily extracted from the full Lagrangian (\ref{fullL}).
It reads
\begin{eqnarray}
{\cal L} & = & \overline\psi (i\, \dslash_4 - k \partial_5 \gamma_5) \psi
+ \overline{\widetilde\psi} (i\, \dslash_4 - k \partial_5 \gamma_5) {\widetilde\psi}
- M (\overline \psi {\widetilde\psi} + \overline{\widetilde\psi} \psi)\nonumber\\
&& + \delta(y) [ \overline q_R i\, \dslash_4 q_R + e \overline q_R \psi_L + e \overline \psi_L q_R],
\label{bulk-boundary}
\end{eqnarray}
where $\psi$ and $\widetilde\psi$ are the singlet components of the periodic bulk fermions
$\Psi^{t}_+$ and $\Psi_+^b$, respectively for $q_R=t_R$ and $q_R=b_R$.
For simplicity of notation, we denote the bulk--to--boundary mixing parameter simply as
$e$ in both cases. Due to the latter and the bulk mass $M$, the equations of motion for $\psi$,
$\widetilde \psi$ and $q_R$ are all coupled with each other. The 4D KK mass eigenstates $\chi^{(n)}_{L,R}(x)$
of eq.~(\ref{bulk-boundary}) will then appear spread between the fields $\psi$, $\widetilde\psi$
and $q_R$. In other words, we expand the various fields as follows:
\begin{equation}
\left\{
\begin{array}{l}
q = \sum_n g_n\, \chi^{(n)}_R\,,\\
\psi_{L,R} = \sum_n f^{(n)}_{L,R}(y)\, \chi^{(n)}_{L,R}\,,\\
{\widetilde\psi}_{L,R} = \sum_n {\widetilde f}^{(n)}_{L,R}(y)\, \chi^{(n)}_{L,R}\,,
\end{array}\right.
\label{fnexp}
\end{equation}
where $f^{(n)}_{L,R}(y)$ and ${\widetilde f}^{(n)}_{L,R}(y)$ are the wave functions along the fifth dimension and $g_n$ are
constants. The $\chi$ fields satisfy the 4-dimensional Dirac equation
\begin{equation}
i\, \dslash_4 \chi^{(n)} = m_n \chi^{(n)}\,,
\end{equation}
where $\chi^{(n)} = \chi^{(n)}_L + \chi^{(n)}_R$, and $m_n$ is the mass of the $n^{\textrm{th}}$
state. By solving the equations of motion, one finds two distinct towers of eigenstates:
an ``unperturbed'' tower with only massive fields and a ``perturbed'' one containing
a right-handed massless field. We will also see how the mass spectra found here with the usual KK 
approach match with the one deduced from eq.~(\ref{WF-terms}) using the holographic approach.

\subsection*{The unperturbed tower}

This tower has no component along the localized fields and is entirely build up with
bulk fields. The wave functions are analytic over the whole covering space and
the left-handed ones vanish at $y=0$. The mass levels are given by
\begin{equation}
m_n = \sqrt{M^2 + \left(\frac{k n}{R}\right)^2}\,,
\label{mnunpert}
\end{equation}
but with $n\geq 1$,  without the $n=0$ mode.
The corresponding wave functions are
\begin{equation}
\left\{
\begin{array}{l}
g_n = 0\,,\\
f^{(n)}_L = 0\,,\\
{\widetilde f}^{(n)}_L =\displaystyle \frac{1}{\sqrt{\pi R}} \sin\left(\frac{n}{R} y\right)\,,
\end{array}
\right.
\qquad
\left\{
\begin{array}{l}
f^{(n)}_R =\displaystyle \frac{M}{m_n \sqrt{\pi R}} \sin\left(\frac{n}{R} y\right)\,,\\
\rule{0pt}{1.75em}{\widetilde f}^{(n)}_R =\displaystyle \frac{n k/R}{m_n \sqrt{\pi R}}
\cos\left(\frac{n}{R} y\right)\,.
\end{array}
\right.
\label{waveunper}
\end{equation}
Although $e$ does not explicitly appear in eqs.~(\ref{mnunpert}) and (\ref{waveunper}), its presence
constrains the left-handed fields to vanish at $y=0$ and thus it indirectly enters in the above expressions.
When $e$ vanishes, indeed, one recovers the usual mode $n=0$.

\subsection*{The perturbed tower}

This tower is distorted by the coupling with the localized fields. Its massive levels
are given by the solutions of the transcendental equation
\begin{equation}
\tan \left(\pi R \frac{\omega_n}{k}\right) = \frac{\epsilon^2}{\pi R k \omega_n}\,,
\label{trascmassform}
\end{equation}
where $\omega_n\equiv\sqrt{(m_n^2-M^2)}$. Notice that the masses
defined by eq.~(\ref{trascmassform}) exactly coincide, for $k=1$, with 
 the zeroes of $Z_q$ in eq.~(\ref{WF-terms}).
The mass equation has a tower of solutions with $m_n>M$, whose wave functions
are\footnote{In eq.~(\ref{eqperturbedtower}) and in eq.~(\ref{eqzeromode})
we report the wave functions for $0 \leq y \leq \pi R$,
the continuation for generic $y$ can be obtained using the properties of the fields
under parity and under translation.}
\begin{equation}\label{eqperturbedtower}
\left\{
\begin{array}{l}
f^{(n)}_L =\displaystyle -\frac{\epsilon m_n g_n}{\sqrt{2 \pi R} k \omega_n}
\frac{\cos\left((y-\pi R) \omega_n/k\right)}{\sin(\pi R \omega_n/k)}\,,\\
\rule{0pt}{1.75em}{\widetilde f}^{(n)}_L = 0\,,
\end{array}
\right.
\hspace{.5em}
\left\{
\begin{array}{l}
f^{(n)}_R =\displaystyle \frac{\epsilon g_n}{\sqrt{2 \pi R} k}
\frac{\sin\left((y-\pi R)\omega_n/k\right)}{\sin(\pi R \omega_n/k)}\,,\\
\rule{0pt}{1.75em}{\widetilde f}^{(n)}_R =\displaystyle -\frac{\epsilon M g_n}{\sqrt{2 \pi R} k \omega_n}
\frac{\cos\left((y-\pi R)\omega_n/k\right)}{\sin(\pi R \omega_n/k)}\,,
\end{array}
\right.
\end{equation}
The constants $g_n$ in eq.~(\ref{fnexp}) are determined by imposing the canonical normalization of the 4D fields.
One gets
\begin{equation}
g_n = \left[1 + \frac{\epsilon^2}{2 \pi R k^2 \omega_n^2}
\left(\frac{\pi R m_n^2}{\sin^2 (\pi R \omega_n/k)}
+ k \frac{M^2 - \omega_n^2}{\omega_n}\cot(\pi R \omega_n/k) \right)\right]^{-1/2}\,.
\label{gndef}
\end{equation}

Notice that some care has to be used in taking the limit $\epsilon\rightarrow 0$ in the above expressions, since
one encounters apparently ill-defined expressions in eqs.~(\ref{eqperturbedtower}) and (\ref{gndef}).

\subsection*{The zero mode}

The zero mode is of particular interest, being identified with a SM field. Its wave function is
\begin{equation}\label{eqzeromode}
\left\{
\begin{array}{l}
f^{(0)}_L =0\,,\\
\rule{0pt}{2.5em}{\widetilde f}^{(0)}_L = 0\,,
\end{array}
\right.
\qquad \qquad
\left\{
\begin{array}{l}
f^{(0)}_R =\displaystyle \frac{\epsilon g_0}{\sqrt{2 \pi R} k}
\frac{\sinh\left((y-\pi R)M/k\right)}{\sinh(\pi R M/k)}\,,\\
\rule{0pt}{1.75em}{\widetilde f}^{(0)}_R =\displaystyle \frac{\epsilon g_0}{\sqrt{2 \pi R} k}
\frac{\cosh\left((y-\pi R)M/k\right)}{\sinh(\pi R M/k)}\,,
\end{array}
\right.
\end{equation}
where again $g_0$ is determined by the normalization conditions and equals
\begin{equation}
g_0 = \left[1 + \frac{\epsilon^2}{\pi R M k} \coth(\pi R M/k)\right]^{-1/2}\,.
\label{g0}
\end{equation}
The constant $g_0$ indicates how much of the zero mode field is composed of the localized field.
When $\epsilon\ll 1$ and $g_0\sim 1$, as expected, the wave function of the massless state is mostly given by the localized field, the bulk components (\ref{eqzeromode}) being proportional to $\epsilon$ and thus small.
On the contrary, for large mixing $\epsilon\gg 1$, one has $g_0\ll 1$ and the localized field is
completely ``dissolved'' into the bulk degrees of freedom. In the latter case, then,
the localization of the $\psi_R$ and ${\widetilde\psi}_R$ components of the zero mode is
essentially determined by the adimensional quantity $\pi R M/k$.
The bulk wave functions at $y = \pi R$ are suppressed by a factor $\exp(-\pi R M/k)$
with respect to their value at $y=0$ so that, independently of $\epsilon$, for large values of $\pi R M/k$
the chiral field is still localized at $y=0$.

Summarizing, the zero mode is localized at $y=0$ for small mixing with the bulk fields
and arbitrary bulk masses, or for large bulk masses and arbitrary mixing.
The requirement of having the correct top mass after EWSB implies a large mixing
and a small bulk mass, giving rise to a delocalized top wave function.
\begin{figure}[t]
\begin{minipage}[t]{0.48\linewidth}
\begin{center}
\includegraphics*[width=\textwidth]{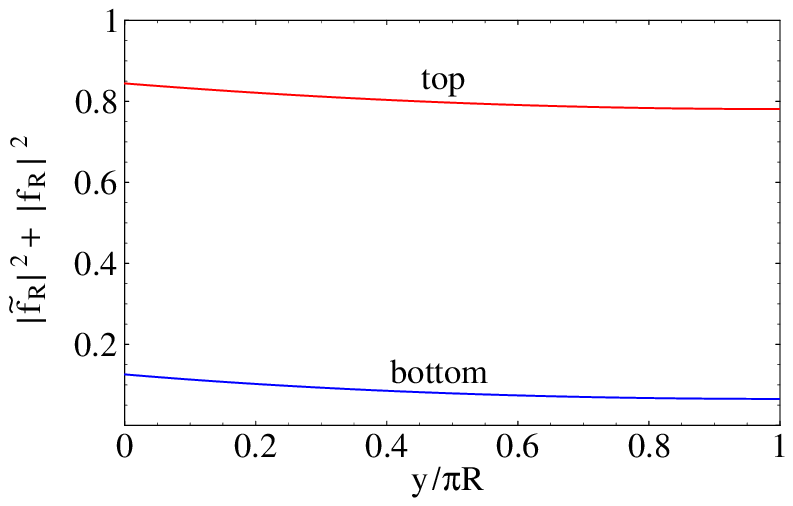}
\caption{Top (red) and bottom (blue) quark wave function profiles (right handed components).
The areas below the lines represent the amount of the delocalized part of the
fields.}\label{figfdotb}
\end{center}
\end{minipage}
\hspace{0.5cm}
\begin{minipage}[t]{0.45\linewidth}
\begin{center}
\includegraphics*[width=\textwidth]{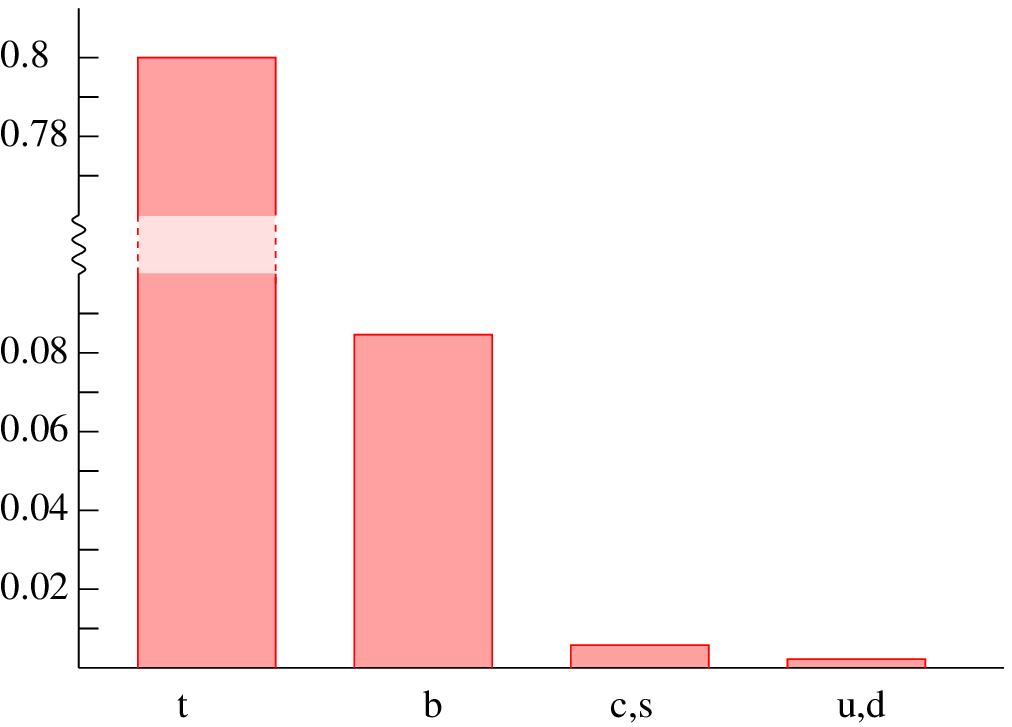}
\caption{Delocalized part of the quark wave functions. The fraction of
bulk wave function is shown for the various quarks.}\label{figlochist}
\end{center}
\end{minipage}
\end{figure}
In fig.~\ref{figfdotb} we illustrate the wave function profiles of the right-handed top and bottom quark
components for typical acceptable values of input parameters, before EWSB.

The mass spectrum and fermion wave functions for
the antiperiodic fermions $\Psi_-$ and $\widetilde\Psi_-$ is straightforward,
since they do  not couple with the localized fields.
Their wave functions can be written as
\begin{equation}
\left\{
\begin{array}{l}
f^{(n)}_L = \displaystyle \frac{1}{\sqrt{2\pi R}} \cos\left(\frac{n+1/2}{R} y\right)\,,\\
{\widetilde f}^{(n)}_L =\displaystyle \frac{\mp 1}{\sqrt{2\pi R}} \sin\left(\frac{n+1/2}{R} y\right)\,,
\end{array}
\right.
\left\{
\begin{array}{l}
f^{(n)}_R =\displaystyle \frac{\mp M-(n+1/2)k/R}{m_n \sqrt{2\pi R}} \sin\left(\frac{n+1/2}{R} y\right)\,,\\
\rule{0pt}{1.75em}{\widetilde f}^{(n)}_R =\displaystyle \frac{M\mp (n+1/2)k/R}{m_n \sqrt{2\pi R}}
\cos\left(\frac{n+1/2}{R} y\right)\,,
\end{array}
\right.
\end{equation}
where $n\geq 0$ and $\pm$ stands for the two towers of mass eigenstates, both with masses given by
\be
m_n= \sqrt{M^2 + \left(\frac{k(n+1/2)}{R}\right)^2}\,.
\ee

After EWSB, all above relations are clearly modified by $O(\alpha)$ effects. In particular,
a small fraction of the SM fields is now spread also among the bulk fermion components
whose mixing were forbidden by $SU(2)_L\times U(1)_Y$ symmetry. We do not write the modification to
the mass formula (\ref{trascmassform}) for the deformed tower, the form of which can be deduced from the
zeroes of the integrand of eqs.~(2.15) and (2.16) of \cite{Panico:2005dh}, with the replacement
$t\leftrightarrow b$.

\end{document}